\documentstyle[11pt,newpasp,twoside,psfig]{article}
\reversemarginpar

\begin{document}
\title{Molecular Gas Distribution in Double-Nucleus Ultraluminous Infrared
Galaxies}
\author{A. S. Evans}
\affil{Dept. of Physics \& Astronomy, SUNY, Stony Brook, NY 11794-3800}
\author{J. A. Surace \& J. M. Mazzarella}
\affil{IPAC, Caltech MS 100-22, Pasadena CA 91125}
\author{and D. B. Sanders}
\affil{Institute for Astronomy, 2680 Woodlawn Dr., Honolulu, HI 96822}

\begin{abstract}

Millimeter (CO) observations of 5 double-nucleus ultraluminous infrared
galaxy (ULIG) mergers are presented.  With nuclear separations of $3-5$
kpc, these galaxies are in the ``intermediate'' stages of the merger
process.  A preliminary comparison of the distribution of molecular
gas (the likely fuel source for both starbursts and active galactic
nuclei: AGN) shows a tendency for molecular gas to be associated with
the AGN nucleus of ULIGs with ``warm'', Seyfert-like infrared colors
($f_{25\mu{\rm m}}/f_{60\mu{\rm m}} \ge 0.20$) and associated with both
stellar nuclei of ULIGs with ``cool'' infrared colors ($f_{25\mu{\rm
m}}/f_{60\mu{\rm m}} < 0.2$). Studies of ULIGs with a wide range of
nuclear separations using the high resolution and increased sensitivity
of ALMA will provide a larger statistical sample with which the gas
distribution, molecular gas masses, and densities can be determined
as a function of the evolutionary stage, starburst and AGN activity,
and lookback time.

\end{abstract}

\section{Introduction}

The availability of ground and space-based instruments
sensitive to radiation redward of 1$\mu$m has resulted in a flurry
of activity in the study of ultraluminous infrared galaxies (ULIG:
$L_{\rm IR} (8-1000\mu{\rm m}) \ge 10^{12}$ L$_\odot$, assuming $H_0$
= 75 km s$^{-1}$ Mpc$^{-1}$).  Much of the present research on these
dusty merger by-products has focussed on two issues - ({\it i})
the nature of the embedded energy sources, and ({\it ii}) the
cosmological significance of dust enshrouded star formation and AGN
activity relative to that observed in optically selected galaxies. The
former issue has been addressed with extensive high-resolution
($0.1\arcsec-0.2\arcsec$), near-infrared imaging with the Hubble Space
Telescope (HST: Scoville et al. 2000), ground-based spectroscopy at
near-infrared wavelengths (e.g. Veilleux, Sanders, \& Kim 1999), and
mid-infrared spectroscopy with the Infrared Space Observatory (Genzel et
al. 1998; Lutz et al. 1998). The data show that at least 30\% of local
ULIGs (i.e., primarily the ``warm'' ULIGs) appear to have AGN as the
dominant energy sources, and that the tendency for the energy source to
be an AGN increases with increasing luminosity. The latter issue has been
investigated via surveys with the Submillimeter Common User Bolometer
Array (SCUBA).  Not only has SCUBA enabled the systematic detection of
ULIGs at cosmological distances (Smail et al.  1997; Hughes et al.
1998; Barger et al. 1998), but compelling arguments infer that ULIGs may
represent a significant fraction of the star formation and AGN activity
occuring in the early universe.

Of equal importance to the question of what ULIGs are is the
question of what their progenitors were. Specifically, it is clear
that ULIGs result primarily from the interaction/merger of gas-rich
spiral galaxies and that they are powered by starbursts and AGN, but
what properties of their progenitor galaxies determine the dominant
energy source, the star formation rate, and the infrared luminosity?
One possible way of answering this is to select a sample of ULIGs
in which the progenitor galaxies are close enough to be under the
gravitational influence of each other, but far enough apart to still
possess distinguishable features. Such a selection criteria has the
benefit of selecting objects which are likely to remain ultraluminous
well into the coalescence of the nuclei, but misses objects that (if
they exist) have similar nuclear separations, but have yet to reach 
the ultraluminous phase. Features which may yield important diagnostics are
the star-forming molecular gas content, the optical or near-infrared
morphologies, and the bulge-to-disk (or perhaps more appropriately,
bulge-to-tidal tail) ratios of the progenitors.

In this conference article, we focus on millimeter (CO) interferometry
of a sample of 5 double-nucleus ULIGs in the redshift range $z \sim
0.05-0.12$ and which have projected nuclear separations of $\sim 3-5$ kpc
($2\arcsec-6\arcsec$). The results presented here were obtained with the
Owens Valley Millimeter Array (OVRO) at a resolution of $\sim 2\arcsec$,
and are discussed at length in Evans et al. (1999), Evans, Surace, \&
Mazzarella (2000), and Evans et al. (2000).

\section{Results and Discussion}

Figure 1 shows the molecular gas distribution of the 5 ULIGs superimposed on 
near-infrared HST images of the galaxies (except Mrk 463, where an
optical HST image is shown). The warm (IR 08572+3915, IR 13451+1232,
and Mrk 463) and cool (IR 12112+0305 and IR 14348-1447) ULIGs are shown
in the left and right panels, respectively. The most striking feature
is the presence of molecular gas on both nuclei of the cool ULIGs,
whereas it appears only on the AGN of the warm ULIGs.

The association of molecular gas with the active nucleus of the
warm galaxies is consistent with the idea that molecular gas is a major
fuel source for their AGN activity. The results are also consistent with
their warm, Seyfert-like mid to far-infrared colors; because H$_2$ 
forms on the surface of dust grains, it is not surprising to find
molecular gas associated with the energy source responsible for heating
the dust to such high temperatures.

In the case of the cool ULIGs, the association of molecular gas with
both of the stellar nuclei (which has also been observed for the double
nuclei in the cool ULIG Arp 220: Sakamoto et al. 1999), combined with
the lack of definitive evidence for AGN (i.e., broad, quasar-like or
high excitation emission lines), may be an indication that nuclear
starbursts in both progenitors enable them to achieve luminosities in
excess of $10^{12}$ L$_\odot$ at this stage of the merger.  There is,
however, no reason to believe twin nuclear starbursts should {\it always}
be necessary to exceed $10^{12}$ L$_\odot$, and thus a larger sample of
ULIGs is required to yield better statistics.

As pointed out by Evans, Surace, and Mazzarella (2000), the morphologies
of the lower luminosity luminous infrared galaxies (LIGs: $L_{\rm
IR} = 10^{11.0-11.99}$ L$_\odot$) with similar nuclear separations
have molecular gas predominantly between the two nuclei. This is a
possible indication that gas stripping is an important process in LIGs,
whereas the bulges of ULIGs may be massive enough such that stripping
is minimal. The starburst and AGN luminosities (and thus the fueling rates)
of ULIGs require sufficient quantities of high density gas, a condition
that can be achieved through dissipative collapse of gas into the deep
nuclear potential wells.  Off-nuclear star formation induced by gas
stripping and compression is unlikely to produce conditions conducive
for the ULIG phenomenon.

\section{ULIGs and the Atacama Large Millimeter Array}

The advantages ALMA will bring to the study of the intermediate stages
of the ULIG process is its ability to perform high resolution observations
with unparalleled sensitivity. ALMA will be able to resolve the molecular
disks of ULIGs with projected nuclear separations as small as $\sim
0.1\arcsec$, and will provide a better constraint on the extent (and thus
the density) of the gas. This is required to determine statistically
whether the properties of molecular gas in more starburst-like ULIGs
differ from that of AGN-like ULIGs, and to compare the gas properties of
ULIGs with LIGs. ALMA will also aid in determining, in a more statistical
sense than possible with present day millimeter arrays, how the general
gas properties of ULIGs vary as a function of lookback time.

\acknowledgements

We thank the OVRO staff and postdoctoral scholars for their assistance
both during and after the acquisition of the data presented here. This
research was supported by NASA grant NAG 5-3042 and RF9736D.

\begin{figure}
\includegraphics{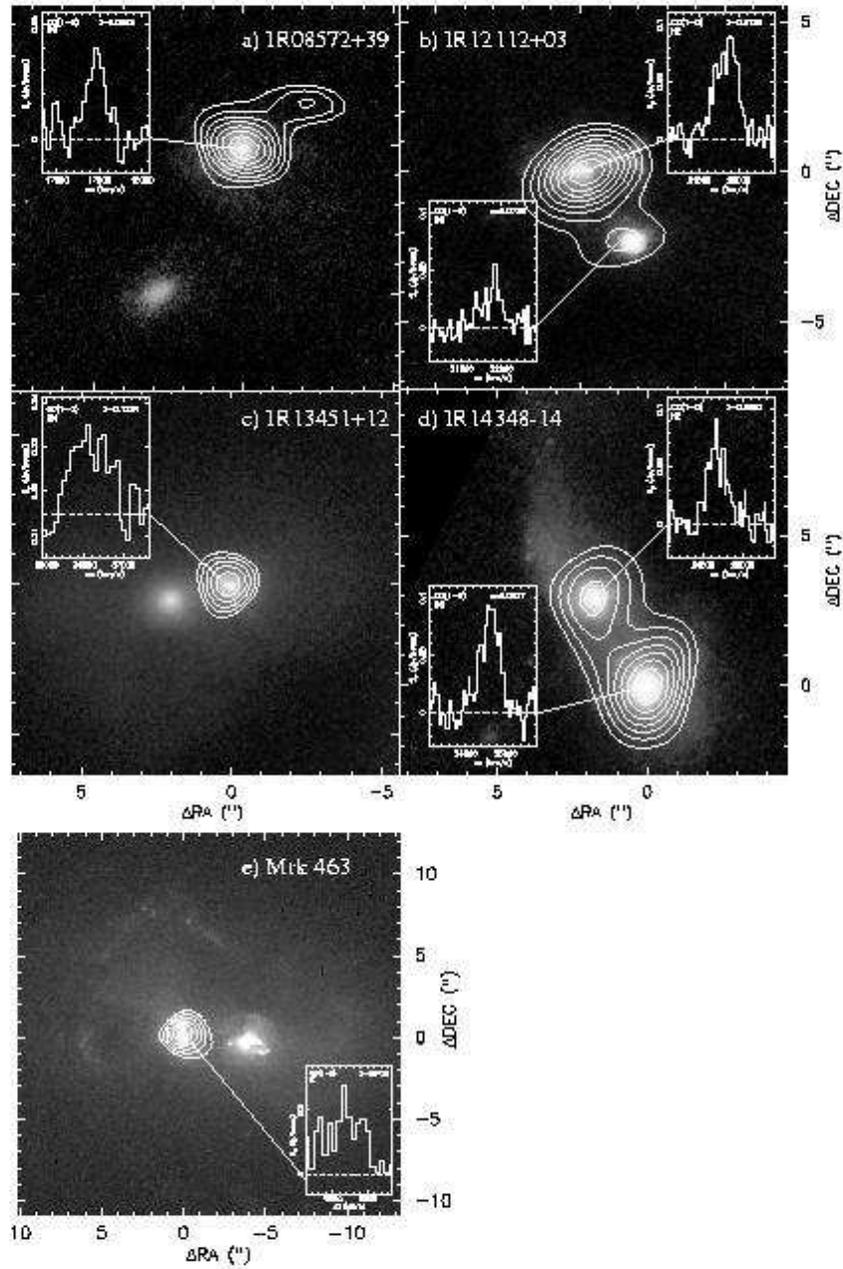}
\vspace{18cm}
\caption{CO($1\to0$) contours of the 3 warm ULIGs (left panels) and the
2 cool ULIGs (right panels) superimposed on HST images.  Extracted CO
spectra of each nucleus are also shown. For each image, north is up and
east is to the left.}

\end{figure}

\end{document}